\documentclass[reprint,nofootinbib,amsmath,amssymb,fontenc aps,longbibliography,preprintnumbers]{revtex4-1}
\usepackage{graphicx,dcolumn,bm,hyperref,float}
\usepackage{anyfontsize}
\usepackage{color}
\usepackage{xcolor}

\newcommand{\beq}{\begin{equation}}
\newcommand{\eeq}{\end{equation}}
\newcommand{\bea}{\begin{eqnarray}}
\newcommand{\eea}{\end{eqnarray}}
\newcommand{\bef}{\begin{figure}}
\newcommand{\eef}{\end{figure}}

\newcommand{\f}{f_a}
\newcommand{\m}{m_a}

\newcommand{\mpl}{M_{\mbox{\tiny{Pl}}}}

\newcommand{\mchi}{m_\chi}

\newcommand{\Y}{Y}

\newcommand{\LamDim}{L}

\newcommand{\mratio}{\mu}
\newcommand{\feff}{f_{\mbox{\tiny{eff}}}}

\newcommand{\cO}{\mathcal{O}}
\newcommand{\cor}{\gamma}
\newcommand{\meff}{m_{\mbox{\tiny{eff}}}}

\begin{document}

\title{Reduced Axion Abundance from an Extended  Symmetry }

\author{Itamar J.~Allali$^1$}
\email{itamar.allali@tufts.edu}
\author{Mark P.~Hertzberg$^{1,2,3,4}$}
\email{mark.hertzberg@tufts.edu}
\author{Yi Lyu$^5$}
\email{ylyu11@ucsc.edu}
\affiliation{$^1$Institute of Cosmology, Department of Physics and Astronomy, Tufts University, Medford, MA 02155, USA
\looseness=-1}
\affiliation{$^2$Center for Astrophysics, Harvard-Smithsonian, Cambridge, MA 02138, USA
\looseness=-1}
\affiliation{$^3$Department of Physics, Harvard University, Cambridge, MA 02138, USA
\looseness=-1}
\affiliation{$^4$Theoretical Physics Center, Department of Physics, Brown University, Providence, RI, 02912, USA
\looseness=-1}
\affiliation{$^5$University of California Santa Cruz, Santa Cruz, CA, 95064, USA
\looseness=-1}

\begin{abstract}
In recent work we showed that the relic dark matter abundance of QCD axions can be altered when the Peccei-Quinn (PQ) field is coupled to very light scalar/s, rendering the effective axion mass dynamical in the early universe. In this work we develop this framework further, by introducing a new extended symmetry group to protect the new particles' mass. We find that with a new global $SO(N)$ symmetry, with large $N$, we can not only account for the lightness of the new scalars, but we can reduce the axion relic abundance in a technically natural way. This opens up the possibility of large PQ scales, including approaching the GUT scale, and still naturally producing the correct relic abundance of axions. Also, in these models the effective PQ scale is relatively small in the very early universe, and so this can help towards alleviating the isocurvature problem from inflation. 
Furthermore, instead of possible over-closure from cosmic strings, the extended symmetry implies the formation of non-topological textures which provide a relatively small abundance.
\end{abstract}

\maketitle

\tableofcontents

\section{Introduction}

The QCD axion continues to be one of the most well-motivated particles beyond the Standard Model (SM). It first emerged as a dynamical solution to the strong CP problem of QCD \cite{Peccei1977}, explaining the smallness of the otherwise allowed $\theta_0 G \tilde{G}$ term in the SM action. This occurs by postulating a new singlet scalar (axion $\theta$) with an approximate shift symmetry. Such a scalar is allowed to couple to the SM by the dimension 5 operator $\theta\, G\tilde{G}$. By carefully studying the ensuing dynamics one finds the effective $\theta$ is driven towards zero. Because such a scalar is electrically neutral and long-lived \cite{Weinberg1978,Wilczek1978}, it has been identified as a viable dark matter (DM) candidate, possibly explaining the missing $\sim85\%$ of the mass of the universe. 

In its basic UV completion, this scalar, axion, arises through the spontaneous breakdown of a new postulated global Peccei-Quinn (PQ) symmetry \cite{Peccei1977} at some high scale $\f$. The axion acquires a small potential and mass from QCD instantons. For generic initial conditions, the axion will be displaced away from the minimum of its potential in the early universe, then later rolling, oscillating, and red-shifting towards zero at late times. This dynamical relaxation towards zero resolves the strong CP problem, while the oscillations themselves behave as nonrelativistic matter in the late universe \cite{Preskill1983,Abbott1983,Dine1983}. In standard cosmologies, the abundance of axions at late times is determined by the symmetry breaking scale, $f_a$ as
\beq
\Omega_a\sim \left(\frac{\f}{ 10^{12}\,\mbox{GeV}}\right)^{7/6}\langle\theta_i^2\rangle
\label{Omaxion}\eeq
with $\langle\theta_i^2\rangle$ denoting the spatial average of the square of the initial displacement angle. For the axion to account for the observed DM ($\Omega_a\approx 0.25$), one needs $\f$ around $2\times10^{11}$ GeV (unless $\langle\theta_i^2\rangle$ is fine-tuned to a small value), giving a mass
\beq
 \m={\Lambda_0^2\over \f}\approx 28 \,\mu\mbox{eV}\left(2\times10^{11}\,\mbox{GeV}\over \f\right)
 \eeq
(where $\Lambda_0\approx 90$\,MeV is related to the QCD scale and quark masses).
In the so-called ``standard axion window," the upper bound on the breaking scale is $f_a\lesssim 10^{12}$\,GeV.
This follows from taking $\langle\theta_i^2\rangle\sim 1$ and then towards the end of this window, the axion will make up all (or most) of the dark matter of the universe, and beyond this upper bound the axion would be too abundant and would over-close the universe. 
 The lower bound of the standard window is $f_a\gtrsim 10^9$\,GeV, from demanding that the axion-photon coupling and axion-nucleon couplings in standard constructions do not produce large effects in stellar physics (e.g. stellar cooling, supernovae cooling, etc.). 

There exist theoretical motivations, however, to prefer values for the PQ symmetry-breaking scale much higher than this standard axion window. It may be more plausible for the physics associated with PQ symmetry breaking to occur at a higher scale suggested by fundamental physics, for instance in grand unified theories (GUT) or string theory \cite{Svrcek2006}.  However, this seems to significantly overproduce the dark matter. There are ways around it by invoking unnaturally small initial misalignment angles $\theta_i$ during inflation; although such models tend to in turn overproduce isocurvature modes in the CMB (see, e.g. \cite{Pi1984,Linde1991,Wilczek2004,Tegmark2006,Fox:2004kb,Hertzberg:2008wr}).

In our previous work \cite{Allali:2022yvx}, we discussed a new framework in which to expand the window of viable $\f$; both to higher and lower values. We proposed a dynamical PQ scale mechanism, wherein the PQ field $\Phi$ is coupled to an additional scalar field $\chi$, which allows the effective PQ scale to evolve with time. This evolution alters the predicted abundance of axions and consequently widens the axion window. However, the lightness of the $\chi$ field was left unexplained.

In this work, we develop this framework further, by addressing the issue of the lightness of the $\chi$ field. We do so by introducing an extended symmetry group, involving the PQ field $\Phi$ and an additional $N_\chi$ scalars $\chi_j$. This new enlarged PQ sector has a total of $N_\chi+2$ scalar degrees of freedom (two from $\Phi$ and $N_\chi$ from the $\chi_j$), which organize into an approximate $SO(N_\chi+2)$ symmetry.
As is standard, QCD instantons explicitly break the $U(1)$ subgroup involving the complex $\Phi$, leaving behind a $SO(N_\chi)$ symmetry. With typical initial conditions, this model has only one free parameter, the number of additional fields $N_\chi$, which determines to which degree the abundance of the axion is suppressed as compared to the standard QCD axion. Finally, there is also the introduction of a mass of $\chi$, which is allowed to be small as it represents a small breaking of the original $SO(N_\chi+2)$ symmetry; so its lightness is technically natural. By making the mass non-zero, the $\chi$ fields eventually relax to the bottom of their potential, and if their mass is small, their abundance is small too. As we will show, for a sufficiently large symmetry group, this scenario can accommodate $\f\gg 10^{12}$ GeV while keeping the axion abundance below the upper bound $\Omega_a\lesssim0.25$ to avoid over-closing the universe. For some other interesting mechanisms that alter the axion's dynamics and abundance, see e.g. \cite{Dvali1995,Heurtier2021,Co2018,Co2019,Co2019a,Co2020,Kitano2021,Nakayama2021,Kobayashi2021,Dienes2016,Dienes2017,DiLuzio2020,Ramazanov2022}.

For such high $\f$, one is normally concerned about the symmetry being broken before the end of inflation, which could generate appreciable isocurvature modes. But in the presence of many fields, the symmetry can be restored, avoiding this problem. 

The outline of our paper is as follows: 
In Section \ref{Std+Dyn} we briefly review the standard QCD axion and our previous dynamical PQ model.
In Section \ref{SON} we introduce the model with an extended symmetry group.
In Section \ref{CosmoEvolution} we discuss the basics of the cosmic evolution. 
In Section \ref{Numerical} we perform a numerical analysis of the equations of motion in the homogenous approximation.
In Section \ref{Discuss} we discuss constraints from isocurvature bounds during inflation, defects, unitarity bounds, and the plausibility of this construction.
Finally, in Appendix \ref{AppAction} we give some more details of the effective action, in Appendix \ref{Inhomogeneities} we describe the eigenmodes from inhomogeneities, and in Appendix \ref{Resonance} we study possible resonance into such inhomogeneities.

\section{Static and Dynamic Peccei-Quinn Mechanisms}\label{Std+Dyn}

\subsection{Static Recap}

The starting point for the canonical Peccei-Quinn (PQ) mechanism involves a complex PQ scalar field $\Phi=\rho\,e^{i\theta}$, which enjoys a global $U(1)_{\mbox{\tiny{PQ}}}$ symmetry. The axion is the angular degree of freedom $\theta$ which becomes a (pseudo-) Goldstone boson when the symmetry is spontaneously broken. The effective Lagrangian density for this axion is given by
\beq
\mathcal{L} = \sqrt{-g}\left[{1\over2}|\partial\Phi|^2-{\lambda\over 4}(|\Phi|^2-\f^2)^2-V(\theta,T)\right]
\label{LStandard}\eeq

Interactions with the Standard Model give the axion a potential, $V(\theta,T)\approx \Lambda(T)^4(1-\cos\theta)$ (in fact there are $\mathcal{O}(1)$ corrections to this shape, but they will not be important here). Consequently the axion acquires a mass, which  depends on temperature. Its low temperature value $m_a$ is related to $\Lambda_0^4=f_a^2m_a^2$, where $f_a$, the PQ scale, is the vacuum expectation value (VEV) of the radial PQ field $\rho$ after symmetry breaking. The choice of the scale $f_a$ for the symmetry breaking dictates the quantity of axions produced by the misalignment mechanism in the early universe according to eq.~\ref{Omaxion}. 

Rather than discussing $\Omega_a$, which is time dependent, it is convenient to use a more fundamental abundance parameter \cite{Hertzberg:2008wr}
\beq
\xi_a(T)\equiv{\rho_a(T)\over n_\gamma(T)} 
\eeq
with $T$ the temperature of the universe. At late times, this tends to a constant as both the numerator and denominator redshift together as $1/a^3$. The observed DM abundance ($\Omega_a\approx0.25$) at late times is the value
\beq
\xi_{obs}\approx 2.9\,\mbox{eV}
\eeq
For further details on the standard axion setup and evolution, see \cite{Allali:2022yvx} and others (e.g.~\cite{DiLuzio2020,DiCortona2016} and references therein).

\subsection{Dynamic Recap}

In our recent work \cite{Allali:2022yvx}, we proposed a mechanism to alter the abundance of axions produced by the misalignment mechanism. We introduced a new scalar degree of freedom $\chi$ which couples to the axion in a way that makes the PQ scale $f_a$ effectively dynamical. This can result in viable abundance predictions for axions outside of the standard allowed window for $f_a$. For instance, in the unaltered misalignment mechanism, one can place an upper bound on $f_a \lesssim 10^{12}\mbox{ GeV}$ to avoid over-closing the universe with too large a density of axions. However, with the increasing-PQ-scale model discussed in \cite{Allali:2022yvx}, we can accommodate $f_a \sim 10^{16} \mbox{ GeV}$, motivated by physics at the GUT scale.

The action of eq.~\ref{LStandard} was modified to be
\bea
\mathcal{L} = \sqrt{-g}\Big{[}{1\over2}|\partial\Phi|^2-{\lambda\over 4}(|\Phi|^2-f(\chi)^2)^2-V(\theta,T)\nonumber\\
+{1\over2}(\partial\chi)^2-{1\over2}\mchi^2\,\chi^2\Big{]}
\label{LDPQ}\eea
where the function $f(\chi)$ determines how the new field $\chi$ couples and thus the time-dependent behavior of the now dynamical PQ scale. As in eq.~\ref{Omaxion}, the energy density of the axions $\rho_a\propto\f^{7/6}$, and it is also proportional to the axion mass. Since the mass is inversely proportional to $\f$, as the effective $\f$ increases with time, the effective $m_a$ decreases, and the abundance is suppressed by an overall factor of $1+7/6$ powers of the effective initial $f_i=f(\chi_i)$
\beq
{\xi_{a}\over\xi_{a,std}}=\left(f_i\over\f\right)^{13/6}
\label{xiax}\eeq
where $\xi_{a,std}$ is the abundance of the standard QCD axion and $\xi_{a}$ is that of the axion in the dynamical PQ scale model.

\section{Extended Symmetry }\label{SON}

Although our work \cite{Allali:2022yvx} had nice phenomenological success, it left the question of the lightness of $\chi$ unexplained. In particular, in the above action there is no symmetry protecting the $\chi$ from being heavy, and so its lightness appears tuned.

\subsection{New Class of Models}

In this work, we wish to stay within this overarching framework, but provide a concrete example in which the mass of $\chi$ is protected by a new global symmetry. It will turn out that in order to alter the axion abundance appreciably, we will need many new scalar fields. Correspondingly, we will need to appeal to an extended symmetry group involving all $N_\chi+2$ degrees of freedom in the new enlarged axion sector ($N_\chi$ new fields and 2 components of the complex PQ field).

We will consider the following updated action, involving $N_\chi$ new scalars $\chi_j$
\bea
\mathcal{L} &=& \sqrt{-g}\bigg{[}{\frac{1}{2}}\Big(|\partial\Phi|^2+\sum_{j=1}^{N_\chi}(\partial\chi_j)^2\Big)\nonumber\\
&&-{\lambda\over 4}\Big(|\Phi|^2+\sum_{j=1}^{N_\chi}\chi_j^2-f_a^2\Big)^2-V(\theta,\chi_j,T)\bigg{]}
\label{LSON}\eea
Apart from the potential $V$, which will be self-consistently taken to be very small, this action is invariant under an $SO(N_\chi+2)$ transformation between the two degrees of freedom of $\Phi=\rho\,e^{i\theta}$ (real and imaginary) and the $N_\chi$ new scalars.

In the very early universe, all of the scalars participate in the $SO(N_\chi+2)$ symmetry. At low energies, the potential term in eq.~\ref{LSON} proportional to $\lambda/4$ spontaneously breaks the symmetry to a residual $SO(N_\chi+1)$ group, leaving $N_\chi+1$ Goldstone bosons. This results in the field $\rho$ (the radial mode of $\Phi$) acquiring a VEV $\langle \rho \rangle = f_i$, while the remaining scalars $\theta$ and $\chi_j$ obtain random initial values $\theta_i$ and $\chi_{j,i}$ and are prevented from evolving by Hubble friction. The initial value of the ``effective" PQ scale
\beq f_{i}^2=f_a^2-\sum_j^{N_\chi}\chi_{j,i}^2< f_a^2
\label{feff}
\eeq
is thus smaller at early times than the vacuum value $f_a$.

\subsection{Induced Masses}\label{InducedMass}

As the temperature of the universe begins to approach the QCD phase transition, QCD instantons induce a potential for the axion, explicitly breaking the $U(1)$ subgroup of the above extended symmetry group, leaving a residual unbroken $SO(N_\chi)$. This leaves $N_\chi$ Goldstone bosons, which we can identify as the $\chi$ particles. With this mass for the axion, it begins to roll down its potential when $3H\sim m_a$, at a temperature $T_{osc}$ as is usual. Its oscillations behave as cold dark matter (CDM). 

Also, the residual $SO(N_\chi)$ symmetry can allow for a mass term for $\chi$. We can write this as
  (note that in eq.~\ref{LSON}, both of these potentials are represented by $V$) 
\beq 
V(\theta,\chi_j,T)\approx\frac{1}{2}m_\chi^2\sum_j^{N_\chi}\chi_j^2+\Lambda(T)^4(1-\cos\theta)
\label{Vchi}\eeq
where $m_\chi$ is the same mass for each $\chi_j$. The presence of this mass, means that eventually the $\chi$ will relax to zero. However, we would like to assume that $m_\chi< m_a(T_{osc})$ such that the axion begins its oscillations first, and the presence of $\chi$ alters in a crucial way the axion evolution. (If the $\chi$ were very heavy, we could just integrate it out, and it would play no important role for the axion). 

Note that since our theory began with an $SO(N_\chi+2)$ symmetry at leading approximation, which prevents a $\chi$ mass, it is technically natural for $\chi$ to be light as it represents a small explicit breaking of this extended symmetry. Since the axion's small mass explicitly breaks the extended symmetry, then so too it will generate a small mass for $\chi$. By considering a one-loop diagram provided by the interaction $\Delta\mathcal{L}=-\sum_j\chi_j^2(\partial\theta)^2/2$ (see Appendix \ref{AppAction} for the relevant action of the low energy theory), one can compute this mass. It can be readily estimated as $m_\chi^2\sim m_a^2\Lambda_{UV}^2/((4\pi)^2f_a^2)$, where $\Lambda_{UV}$ is a cutoff on the loop integral; this should be taken to be of the order the mass of the radial PQ mode $\Lambda_{UV}\sim m_{PQ}=\sqrt{2\lambda}\,f_a$. So the induced mass is $m_\chi^2\sim \lambda\,m_a^2/(4\pi)^2$. As discussed in Section \ref{Unitarity} we already know that we need to take $\lambda$ somewhat small to maintain perturbative unitarity, so this self-consistently implies $m_\chi\ll m_a$, as we assume in this work.

The above QCD axion potential is taken to be of the standard form $\Delta V = \Lambda(T)^4(1-\cos\theta)$. However, we note that this should not be used when the magnitude of the PQ field $\rho$ happens to be very small or vanish, since in that regime $\theta$ is not well defined. In this paper, we will generally assume that $\rho$ is not particularly small. In fact as we will explain in the next section, the initial condition is naturally on the order $\sqrt{2}\,f_a/\sqrt{N_\chi+2}$ and its late time value is $f_a$. 
However, one can consider the case in which $\rho$ is accidentally much smaller. Here one should alter the potential $\Delta V$ accordingly. A parameterization of an anticipated potential that incorporates both the angular dependence and the radial dependence is of the form $\Delta V = (\rho^2/(M^2+\rho^2))\Lambda(T)^4(1-\cos\theta)$. In this parameterization, when $\rho\to 0$ then indeed $\Delta V\to0$, and the dependence on $\theta$ disappears. On the other hand, for large values of $\rho\gg M$ the effective potential becomes non-zero and the generation of the potential for $\theta$ from QCD instantons becomes standard. Hence as long as $M$, the characteristic cross-over scale, is somewhat smaller than $f_a$ (depending on $N_\chi$), then one expects our upcoming primary results to be unaltered. 

We also note that the construction in this paper has some overlap with the very interesting Ref.~\cite{Chaumet:2021gaz}. As is seen in that paper, the final abundance is reduced for many fields; we shall find compatible results here. 

An important question is whether this all has an embedding within a UV complete model, with heavy fermions, etc. (requiring a significant extension beyond the minimal models \cite{Kim1979,Shifman1980,Zhitnis1980,Dine1981}); but we leave this for future consideration.

\section{Cosmological Evolution}\label{CosmoEvolution}

After the spontaneous breaking of the extended symmetry group (occurring well before the QCD phase transition since $\f$ is very large), the value of $\rho$ is frozen as
\beq
\rho=\sqrt{f_a^2-\sum_{j=1}^{N_\chi}\chi_j^2}
\eeq
Using $|\Phi|=\rho$ and $|\partial\Phi|^2=(\partial\rho)^2+\rho^2(\partial\theta)^2$, we insert this into the above action and
obtain a kind of nonlinear $\sigma$ model for the remaining $N_\chi+1$ light degrees of freedom; see Appendix \ref{AppAction} for this action.

It is straightforward to vary the action and obtain its classical equations of motion (since these light fields are typically at very high occupancy, the classical field approximation should suffice here).
To write down the equations of motion, we work with dimensionless variables defined by 
\beq
\tau\equiv\m\,t,\,\,\,\,\Y_j\equiv{\chi_j\over f_a},\,\,\,\,\LamDim(T)\equiv{\Lambda(T)^4\over\Lambda_0^4},\,\,\,\,\mratio\equiv{\mchi\over\m}
\eeq
In the equations of motion,  we can, to first approximation, ignore spatial variations and focus on the zero modes of the fields. 
A study of inhomogeneities in the fields is given in Appendices \ref{Inhomogeneities} and \ref{Resonance}, where we check for possible instabilities or resonance in the system. Also, a discussion of defects is given in Section \ref{Defects}.

It suffices to ignore anharmonicity of the potential and write $\sin\theta\approx\theta$. Also, we write $H=1/(2\,t)$ in the radiation era treated in the FRW approximation.  The temperature dependence of the axion potential can be estimated as $L(T)\sim (T_{qcd}/T)^8$ for $T\gg T_{qcd}$ and $L(T)\approx 1$ for $T\ll T_{qcd}$ where $T_{qcd}\sim 100$\,MeV is of the order of the temperature of the QCD phase transition. 
While this temperature dependence has been confirmed by recent lattice studies, the $n=8$ exponent is only approximate and may take a slightly different value.

In general there are $N_\chi$ independent new fields. However, due to the residual $SO(N_\chi)$ symmetry, they all evolve in a similar way, only possibly differing by their initial conditions. For simplicity, here we mention the case in which their initial conditions are all equal, giving a set of identical equations of motion $Y\equiv Y_1=Y_2=\ldots Y_{N_\chi}$ as
 \bea
&& \theta_{\tau\tau}+\left({3\over2\,\tau}- {2N_\chi Y Y_\tau\over 1- N_\chi Y^2}\right)\!\theta_\tau
 +\frac{\LamDim(T)}{1-N_\chi Y^2}\theta=0\,\,\,\, \label{EOMtheta}\\
&& \Y_{\tau\tau} +\left({3\over2\,\tau}+{N_\chi YY_\tau\over 1-N_\chi Y^2} \right)\!Y_\tau\nonumber\\
&&\hspace{2.5cm}+ (1-N_\chi Y^2)(\mu^2+\theta_\tau^2) Y =0
  \label{EOMY}
 \eea
 where each subscript $\tau$ corresponds to a derivative in $\tau$ (i.e., $\theta_\tau\equiv d\theta/d\tau,\,Y_\tau\equiv dY/d\tau$, etc). 
 At early times, the scalars are Hubble friction dominated, so we pick initial conditions for their velocities to be $\theta_{\tau,i}=Y_{\tau,i}=0$, which is consistent with the underlying $SO(N_\chi)$ symmetry and the discrete $Y_{j}\to-Y_{j}$ symmetry.
 
 The extension to random initials conditions is analytically simple, though numerically much more complicated. So we shall use this special case, which we believe suffices to illustrate the essential behavior. Nevertheless, it would be useful to extend our analysis to more generic initial conditions. 

\subsection{Initial Conditions}

For the canonical axion, involving only the complex field $\Phi$, the initial condition for $\theta$ is typically taken to be $\cO(1)$. In the scenario when PQ symmetry is broken before the end of inflation, this can be understood as a random typical angle between $[-\pi,\pi]$. In the case of PQ symmetry breaking after the end of inflation, our observable universe is a huge collection of regions acquiring different $\theta$ values at symmetry breaking; this necessarily means that the average of $\langle\theta_i^2\rangle=\pi^2/3$ is $\mathcal{O}(1)$.

With the new extended symmetry group, one anticipates that the symmetry is broken in a random way; the $N_\chi+2$ degrees of freedom each have an equal chance of taking on some value, but the sum of their squares is constrained by the symmetry breaking potential. Let us briefly recast the complex PQ field as $\Phi=\phi_1+i\phi_2$, such that the symmetry breaking results in the constraint
\beq 
\phi_{1}^2+\phi_{2}^2 +\sum_j^{N_\chi}\chi_{j}^2 = f_a^2 \label{constraint}
\eeq
The symmetry ensures that each of the $N_\chi+2$ scalars takes on a random initial value with equal probability distribution. The unconditional probability distribution for each individual random variable $\psi$ (where $\psi$ can be $\phi_{1,i}$ or $\phi_{2,i}$ or $\chi_{j,i}$) can be shown to be
\beq
p(\psi)={1\over\mathcal{M}}(f_a^2-\psi^2)^{(N_\chi-1)/2}
\eeq
where $\mathcal{M}$ is a normalization factor; whose value is $\mathcal{M}=\sqrt{\pi}\,\Gamma((1+N_\chi)/2)f_a^{N_\chi}\!/\Gamma(1+N_\chi/2)$. For large $N_\chi$ this becomes a Gaussian distribution with vanishing mean. Its variance is (which can also be inferred from eq.~(\ref{constraint}))
\beq
\langle\psi^2\rangle=\langle\phi_{1,i}^2\rangle=\langle\phi_{2,i}^2\rangle=\langle\chi_{j,i}^2\rangle={f_a^2\over N_\chi+2}
\eeq 
We choose the
initial conditions for each field to be exactly the root mean square for simplicity. This is
equivalent to fixing $\rho_i^2=\phi_{1,i}^2+\phi_{2,i}^2=2f_a^2/(N_\chi+2)$, and taking $\theta_i = \pi/4$ and $Y_{j,i}=1/\sqrt{N_\chi+2}$. Then, as the $\chi_j$ evolve once Hubble drops below their mass, the effective VEV of $\rho$ ($=\sqrt{\phi_1^2+\phi_2^2}$) shifts according to eq.~\ref{constraint}.

 One may have noticed that the equations of motion eq.s~\ref{EOMtheta} and~\ref{EOMY} may be rendered independent of $N_\chi$ under the change of variables $\tilde{Y}\equiv \sqrt{N_\chi} \,Y$. Then, for $N_\chi \gg 1$, the initial conditions become $\tilde{Y}_{j,i}=\sqrt{N_\chi}/\sqrt{N_\chi+2}\approx 1$ and the parameter $N_\chi$ naively appears to drop out of the model altogether. However, to properly analyze the large $N_\chi$ regime, one should Taylor expand $\tilde{Y}_{j,i}= (1+2/N_\chi)^{-1/2}\approx (1-1/N_\chi)$, meaning terms involving $(1-N_\chi Y^2)$ are initially $(1-\tilde{Y}_{j,i}^2)\approx 2/N_\chi$, which is clearly sensitive to $N_\chi$. 
 
 The key phases of the evolution are as follows: first, the effective friction terms proportional to $\theta_\tau$ in eq.~\ref{EOMtheta} and $Y_\tau$ in eq.~\ref{EOMY}, respectively, become canonical (i.e. $3/2\tau = 3H$) at early times. Next, the effective mass term for the axion is approximately $N_\chi\LamDim(\tau)/2$ which becomes very large with large $N_\chi$; this implies an earlier onset of oscillations and suppressed abundance of axions. Further, the effective mass of $Y$ becomes proportional to $2/N_\chi$, and thus shrinks with growing $N_\chi$, delaying the onset of oscillations for the $Y$ field with larger $N_\chi$. This predicted behavior from examining the equations of motion is verified in the numerical solutions presented in fig.s~\ref{fig:N1}-~\ref{fig:N1000}.

\subsection{Analytical Estimates for Relic Abundance}

\begin{figure}[t!]
    \centering
    \includegraphics[width=\linewidth]{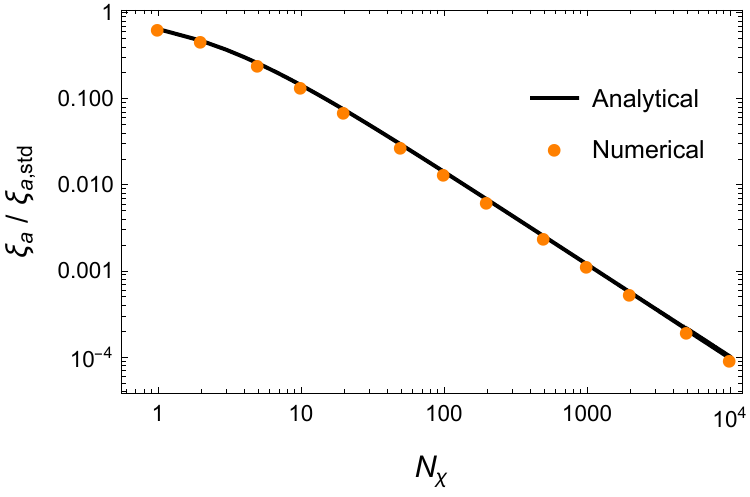}
    \caption{ The numerically computed ratio $\xi_a/\xi_{a,std}$ is shown for various values of $N_\chi$ in the orange points, indicating that the abundance is suppressed with growing $N_\chi$. It can be seen that the analytical prediction (solid black) for the abundance suppression is accurately reproduced by the numerical solution. In this plot, the numerical solutions correspond to $f_a\sim10^{14}$ GeV, but the behavior is expected to be the same for any large $\f$.}
    \label{fig:xi_of_N}
\end{figure}

These initial conditions result in the initial effective PQ scale from eq.~\ref{feff} to be
\beq f_i = 
f_a \bigg(\frac{2}{N_\chi+2}\bigg)^{1/2}
\label{feffNchi}
\eeq
Making the same estimate as in eq.~\ref{xiax}, we make the following analytical prediction $\xi_{a,an}$ for the suppression of the axion abundance at late times
\beq
{\xi_{a,an}\over\xi_{a,std}}=\left(f_i\over\f\right)^{13/6}=\bigg(\frac{2}{N_\chi+2}\bigg)^{13/12}
\label{xiaxNchi}\eeq
Thus, the fractional change in the axion abundance compared to the standard theory should depend solely on the number of additional fields $N_\chi$.

The relic abundance of $\chi$ is more involved to estimate. However, as we showed in our previous paper \cite{Allali:2022yvx}, the case of a single $\chi$ field has an estimated relic abundance $\xi_{\chi,an}$ of
\beq
{\xi_{\chi,an}\over\xi_{a,std}} = {g_\chi^{3/4}\,g_{s*}\over g_*^{7/12}\,g_{s\chi}}{T_{QCD}^{2/3}\over \Lambda_0^{2/3}}{\f^{1/3}\over\mpl^{1/3}}
{\sqrt{\mu}\,\langle Y_i^2\rangle\over\langle\theta_i^2\rangle}
\eeq
(where we have taken eq.~(41) of Ref.~\cite{Allali:2022yvx} and replaced $F\to 1$, as is appropriate to match onto the model here). Now in order to account for the total energy density stored in all of our $N_\chi$ fields, we simply sum over $N_\chi$. And to account for our initial conditions, we choose $\langle Y_i^2\rangle =1/(N_\chi+2)$. Hence the total is
\beq
{\xi_{\chi,an,total}\over\xi_{a,std}} = {g_\chi^{3/4}\,g_{s*}\over g_*^{7/12}\,g_{s\chi}}{T_{QCD}^{2/3}\over \Lambda_0^{2/3}}{\f^{1/3}\over\mpl^{1/3}}
{\sqrt{\mu}\over\langle\theta_i^2\rangle}{N_\chi\over N_\chi+2}
\eeq
Interestingly, this shows that for large $N_\chi$, the abundance is fixed. Hence in order to ensure the $\chi$ abundance is small, we only need to assume a small $\chi$ mass, which suppresses this through the factor $\sqrt{\mu}=\sqrt{m_\chi/m_a}$. Since the $\chi$ mass is protected by our original symmetry, a small $\chi$ abundance is plausible within the effective theory.

We note that this simple analytical estimate for $\xi_\chi$ tends to overpredict the abundance of $\chi$. This is because the estimate relies on the mass term $\mu^2$ in eq.~\ref{EOMY} controlling the onset of oscillations of $\chi$. However, in the regime when $\theta_\tau^2>\mu^2$, this hierarchy causes the $\chi$ field to oscillate a bit earlier, and therefore have a smaller abundance than predicted here. These formulae for $\xi_\chi$ become more accurate when $\theta_\tau^2 < \mu^2$ when $H\sim m_{\chi,\mbox{\tiny{eff}}}$. Crucially, this ambiguity does not influence the prediction for the abundance of the axion because, regardless of the time of $\chi$ oscillations, the axion always oscillates first and thus its abundance is determined appropriately.

\section{Numerical Analysis}\label{Numerical}

\begin{figure}[t!]
    \centering
    \includegraphics[width=\linewidth]{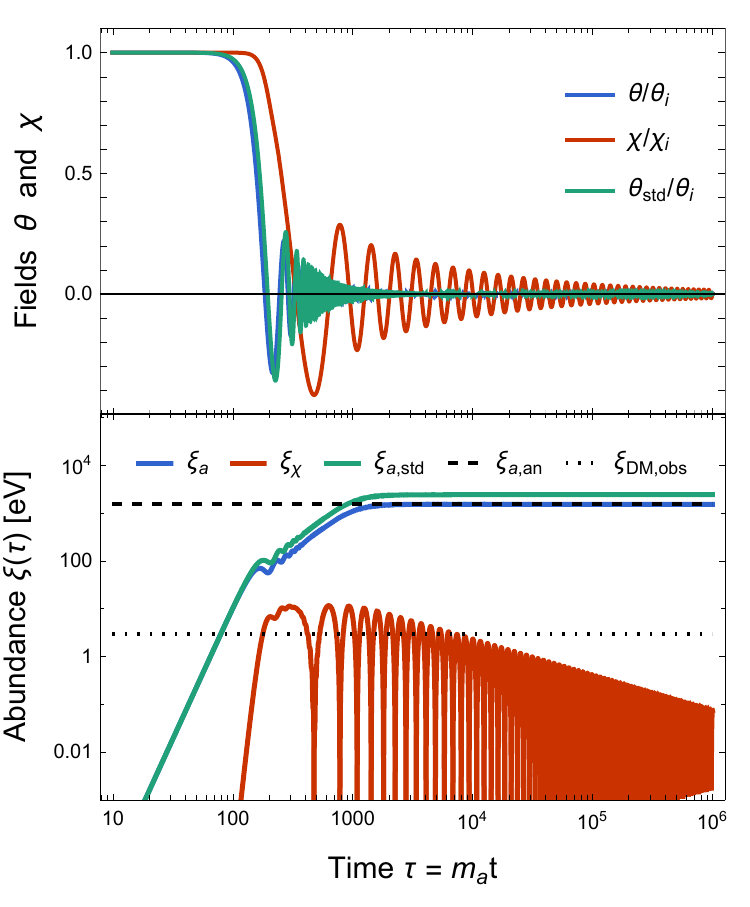}
    \caption{The results of numerical solutions of the field equations are shown for $\f\sim 10^{14}$ GeV with $N_\chi = 1$ and $m_\chi/ m_a = 10^{-5}$. Top: The time evolution of the $\theta$ and $\chi_j$ fields.
    Bottom: The abundances of $\theta$ and $\chi_j$.}
    \label{fig:N1}
\end{figure}

\begin{figure}[t!]
    \centering
    \includegraphics[width=\linewidth]{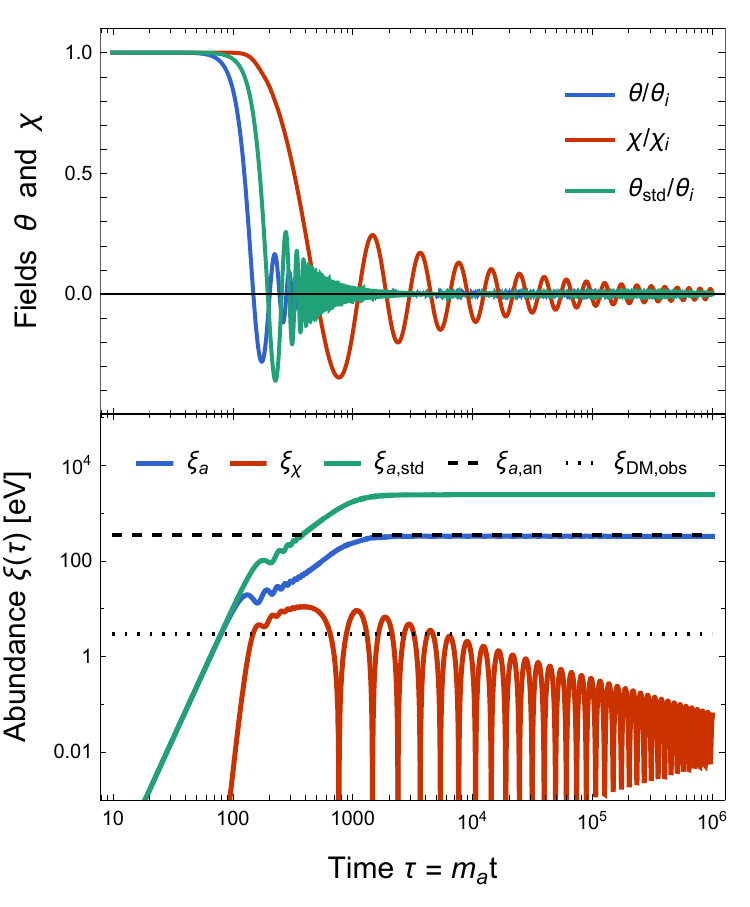}
     \caption{The results of numerical solutions of the field equations are shown for $\f\sim 10^{14}$ GeV with $N_\chi = 10$ and $m_\chi/ m_a = 10^{-5}$. Top: The time evolution of the $\theta$ and $\chi_j$ fields.
    Bottom: The abundances of $\theta$ and $\chi_j$.}
    \label{fig:N10}
\end{figure}

\begin{figure}[t!]
    \centering
    \includegraphics[width=\linewidth]{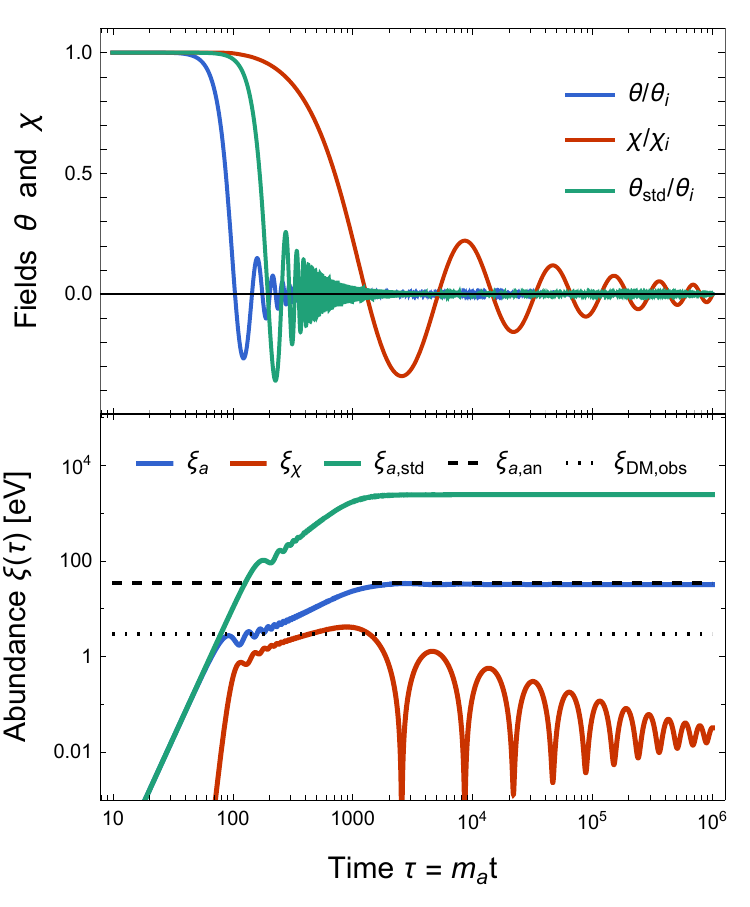}
     \caption{The results of numerical solutions of the field equations are shown for $\f\sim 10^{14}$ GeV with $N_\chi = 100$ and $m_\chi/ m_a = 10^{-5}$. Top: The time evolution of the $\theta$ and $\chi_j$ fields.
    Bottom: The abundances of $\theta$ and $\chi_j$.}
    \label{fig:N100}
\end{figure}

\begin{figure}[t!]
    \centering
    \includegraphics[width=\linewidth]{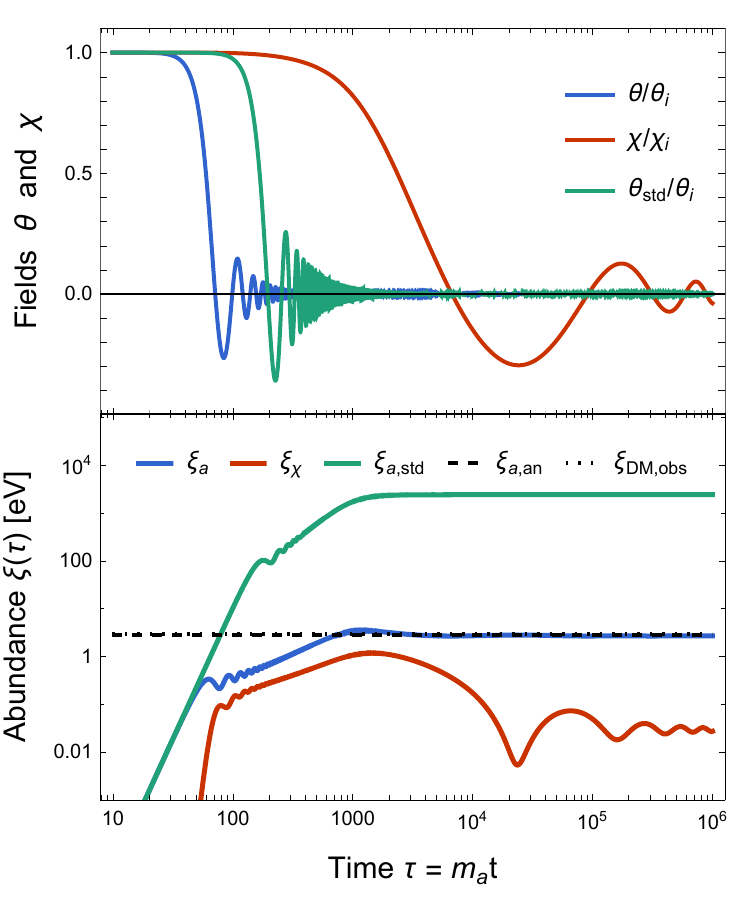}
     \caption{The results of numerical solutions of the field equations are shown for $\f\sim 10^{14}$ GeV with $N_\chi = 1000$ and $m_\chi/ m_a = 10^{-5}$. Top: The time evolution of the $\theta$ and $\chi_j$ fields.
    Bottom: The abundances of $\theta$ and $\chi_j$.}
    \label{fig:N1000}
\end{figure}

The dynamics of the model can be solved for numerically. Recall that the $N_\chi$ additional fields are treated as having identical histories, and therefore one needs only to solve the two coupled differential equations~\ref{EOMtheta} and~\ref{EOMY}. Below, we present the time-evolution and the effects of choosing different numbers $N_\chi$. We show first the behavior of the altered abundance compared to a standard axion $\xi_a/\xi_{a,std}$ in fig.~\ref{fig:xi_of_N}. We see good agreement between our analytical prediction for the axion relic abundance (solid line) and the numerically obtained abundance.

Next, we display the detailed numerical results for the case $f_a\sim 10^{14}$ GeV as an example, choosing $N_\chi=1,\,10,\,100,\,1000$ in fig.s~\ref{fig:N1}-\ref{fig:N1000}. We see in these examples too that the analytical estimate (dashed line) matches well with the numerical solution.

To raise towards simple estimates for the grand unified scale, one could strive for $f_a\sim10^{15}-10^{16}$ GeV.
However, numerically solving for the dynamics of these choices for large $N_\chi$ presents challenges. To capture the behavior of the light $\chi_j$ fields, one needs to integrate through large amounts of time, at which point the rapid oscillations of the axion become difficult to handle numerically. Though we are confident that the results presented for $f_a\sim 10^{14}$ GeV can be used to infer results for larger $f_a$, it remains important to verify this explicitly with more sophisticated numerical methods. Furthermore, the effects of inhomogeneities may be important. We discuss these effects in Appendix~\ref{Inhomogeneities}, arguing that the homogeneous analysis is sufficient to capture the behavior of the model.

In any case, our existing numerics shows that for large $N_\chi$, the final axion abundance is reduced as anticipated. Extending to larger $f_a$, we need larger $N_\chi$; $f_a\sim10^{15}$ and $10^{16}$ GeV require $N_\chi\sim 10^{4}$ and $10^5$ new species, respectively, with corresponding huge symmetry groups.

\section{Discussion}\label{Discuss}

There are many interesting points to discuss within this framework. We shall discuss several key points in this section.

\subsection{Isocurvature Fluctuations}

As is well known, if there are light fields present during cosmic inflation, they acquire a de Sitter temperature and fluctuate as $\delta\phi\sim T_{dS}=H_{\mbox{\tiny{inf}}}/(2\pi)$ per Hubble time. If such fields go on to provide a significant fraction of the dark matter, then this translates into significant isocurvature fluctuations at early times, which leave an imprint on the CMB (e.g., see refs.~\cite{Fox:2004kb,Hertzberg:2008wr}.) Then if the Hubble scale of inflation is large, the amplitude of these isocurvature modes is large, and ruled out by current constraints. 

In standard axion models with very high $\f$, the PQ symmetry is spontaneously broken during inflation, as $T_{dS}=H_{\mbox{\tiny{inf}}}/(2\pi)\ll \f$. This results in the light axion forming during inflation and giving rise to large isocurvature fluctuations. However, in the presence of our new class of models the situation is altered. In the presence of many fields, the condition for symmetry breakdown depends on the combined fluctuations of our $N_\chi+2$ fields 
\beq
\Delta^2=(\delta\phi_1)^2+(\delta\phi_2)^2+\sum_{j=1}^{N_\chi}(\delta\chi_j)^2
\eeq
Symmetry breakdown occurs when these fluctuations are smaller than the PQ sale $f_a$, i.e., $\Delta\lesssim f_a$. Since each field acquires the de Sitter fluctuations $\delta\phi\sim T_{dS}$, the condition is
\beq
T_{dS}\lesssim {\f\over\sqrt{N_\chi+2}}
\label{isocond}
\eeq
This condition is much more difficult to satisfy when $N_\chi$ is large, as we are assuming here. Therefore it is much more plausible that the symmetry remains unbroken during inflation. Currently inflation has its Hubble scale bounded by the lack of detection of B-modes in the CMB. This corresponds to a bound on the tensor to scalar ratio of $r\lesssim 0.04$ \cite{Ade2021,Akrami2020}. Converting this to a Hubble scale and in turn a de Sitter temperature, we have $T_{dS}\lesssim 10^{13}$\,GeV. Hence even if we push $\f$ towards $\sim 10^{14}$\,GeV, we can plausibly violate inequality~\ref{isocond} with  $N_\chi\sim 10^3$. In this case, there are no appreciable isocurvature modes generated, which is compatible with current data. Then, pushing to even higher $\f$ requires larger $N_\chi$ to suppress its abundance, and this larger $N_\chi$ in turn alters the condition in eq.~\ref{isocond}. For $\f\sim 10^{15}$ GeV, the required $N_\chi\sim10^{4}$ is sufficient to obtain symmetry-breaking after inflation, while for $\f\sim10^{16}$ GeV, one needs more than the required $N_\chi\sim10^{5}$ to avoid isocurvature bounds.

On the other hand, as the scale of inflation (and the de Sitter temperature) are lowered, the condition for symmetry breakdown is easier to obey and appreciable isocurvature modes can arise. We note that in our setup here, this is primarily carried by the axion, as the abundance of the $\chi$ fields are small.
Other ideas to suppress the isocurvature modes include \cite{Bao2022}.

\subsection{Defects}\label{Defects}

When spontaneous symmetry-breaking occurs after the end of inflation, as is argued in the previous section, there can be the creation of topological defects. For the standard PQ models, cosmic strings arise from the Kibble mechanism where super horizon regions of space acquire a different $\langle \theta_i\rangle$, connecting in configurations with nontrivial winding. When the $U(1)_{PQ}$ is explicitly broken by QCD instantons, the number of degenerate vacua in the periodicity of $\theta \in [0,2\pi)$ dictate the number of domain walls attached to each string ($N_{DW}$) in the string-domain-wall network of topological defects. 

If $N_{DW}>1$, the network is stable and dominates the universe, over-closing it. If $N_{DW}=1$, there are no stable domain walls, only cosmic strings. After the QCD phase transition, the would-be domain walls are pulled together and decay, producing an additional source of axions.  It has been debated in the literature whether this enhances the final abundance compared to the misalignment value by a factor of a few or dozens (see e.g., Refs.~\cite{Hindmarsh:2019csc,Gorghetto:2020qws}). Regardless, the standard axion requires the consideration of this topological defect network to accurately predict the axion abundance.

In contrast, when the classical symmetry of the scalar sector has been promoted from $U(1)=SO(2)$ to $SO(N_\chi+2)$, the story is quite different. When the spontaneous symmetry breaking occurs, a string network will not form via the Kibble mechanism, since there is now multiple Goldstone bosons. For example, if $N_\chi=1$, the spontaneously broken $SO(3)$ leaves behind two Goldstones, leading to the creation of monopoles. For larger $N_\chi$, textures can be formed. For $N_\chi\ge 3$ these are non-topological and collapse when entering the horizon and so they have a small relic abundance. Outside the horizon, they enter a kind of scaling solution for large $N_\chi$ \cite{Spergel:1990ee,Turok:1991qq,Hu:1996yt,Amin:2014ada}. Any residual imprints from textures, would be a signature of this construction. However, the presence of the axion and $\chi$ masses will suppress textures at later times by making $\theta=\chi_j=0$ a preferred value. In any case, for $N_{DW}=1$, the theory is safe from over-closure from defects.

For $N_{DW}>1$, there are still potentially dangerous domain walls from the axion's $N_{DW}$ discrete minima. However, now the remaining $SO(N_\chi)$ symmetry, which is unbroken by QCD instantons, enhances the vacuum manifold to a $N_\chi$-sphere. This prevents the stability of such domain walls. This is because even if locally the axion is trapped in one of its discrete minima, it can shed energy into this degenerate $N_\chi$-sphere until it reaches another discrete vacuum, removing a domain wall. However, eventually the small but non-zero $\chi$ mass becomes relevant and suppresses this process. A full analysis of this process is beyond the scope of the current work.

\subsection{Unitarity Bounds}\label{Unitarity}

A concern in the model is that with a very large number of scalars, one should check that the theory remains unitary. Our fields have quartic interactions with one another of the form
\beq
\Delta\mathcal{L}=-\sum_{i,j}{\lambda\over 4}\chi_i^2\chi_j^2
\eeq
So for example, if we compute the annihilation cross section of a pair of particles ``1" into any final state at energies above the PQ radial mass $m_{PQ}=\sqrt{2\lambda}\,f_a$, we have
\beq
\sigma_{1\to \mbox{\tiny{All}}}\approx{\lambda^2(N_\chi+10)\over 8\pi\,E_{cm}^2}
\eeq
For large $N_{\chi}$ we risk violating the unitarity bound $\sigma\leq (2\pi)/E_{cm}^2$. Thus, we need to scale down $\lambda\leq(4\pi)/\sqrt{N_\chi+10}$ to avoid this problem. So if $N_\chi\sim 10^4$, we need to impose $\lambda< 10^{-1}$, or so, to maintain perturbative unitarity. However, this does not appear a huge problem. In fact it self-consistently reinforces the lightness of $\chi$, as discussed in Section \ref{InducedMass}.

\subsection{Future Work and Plausibility}

A very important question for future consideration is the plausibility of this new (large) symmetry group. Ideas within unification often involve large groups, such as $SO(10)$, etc, but we are making a case for potentially even much larger groups. Can this fit in and improve ideas within unification, or does it make the situation more difficult?

Relatedly, it is important to develop  microscopic models with fermions. In standard axion models, there are additional heavy fermions that are charged under the $U(1)_{PQ}$ symmetry. Naively this breaks our starting $SO(N_\chi+2)$ symmetry already. So it remains an open question to develop alternate models with fermions that may account for this altered symmetry structure. As it stands, we have a consistent effective field theory for a collection of scalars, one of which -- the axion -- is assumed to couple to gluons with a dimension 5 coupling $\theta\,G\tilde{G}$; QCD instantons still generate a potential for this and it still solves the Strong CP problem, as the standard axion models do. A full UV completion though is an important direction for future work.

\section*{Acknowledgments}
We thank Jose Blanco-Pillado for helpful comments. 
M.~P.~H and Y.~L acknowledge support from the Tufts Visiting and Early Research Scholars’ Experiences Program (VERSE).
M.~P.~H is supported in part by National Science Foundation grant PHY-2013953. I.~J.~A. acknowledges support from the John F. Burlingame Graduate Fellowship in Physics at Tufts.

\appendix

\vspace{0.3cm}
\section{Reduced Action}\label{AppAction}

After eliminating the radial mode $\rho=|\Phi|$, we obtain a reduced action for the remaining $N_\chi+1$ light degrees of freedom. Let us organize the $\chi_j$ into a vector $\vec{\chi}$ to express it. We obtain
\bea
&&\mathcal{L}=\sqrt{-g}\Bigg{[}{1\over 2}|\partial\vec{\chi}|^2+{1\over2}{(\vec{\chi}\cdot\partial{\vec\chi})^2\over f_a^2-|{\vec\chi}|^2}
+{1\over2}(f_a^2-|{\vec\chi}|^2)(\partial\theta)^2\nonumber\\
&&\hspace{2.8cm}-\,\Lambda(T)^4(1-\cos\theta)-{1\over2}m_\chi^2|{\vec\chi}|^2\Bigg{]}
\eea
From this low energy action, all the results of the paper can be derived. 

\section{Inhomogeneities}\label{Inhomogeneities}

Let us expand around our homogeneous background fields as
\beq
\theta=\theta_0(t)+\varepsilon(t,{\bf x}),\,\,\,\,\chi_j=\chi_0(t)+\gamma_j(t,{\bf x})
\eeq
We wish to work to quadratic order in the action. The zeroth order terms were already solved for numerically in this paper; the first order terms will vanish by the Euler-Lagrange equations. The second order Lagrangian for the perturbations is found to be
\bea
&&{\mathcal{L}\over a^3}={1\over2}|\partial{\vec\gamma}|^2+{\chi_0^2(\sum_j\partial\gamma_j)^2\over2(f_a^2-N_\chi\chi_0^2)}+{1\over2}(f_a^2-N_\chi\chi_0^2)(\partial\varepsilon)^2
\nonumber\\
&&-2\chi_0\dot{\theta}_0\sum_j\gamma_j\,\dot\varepsilon-{1\over2}(m_\chi^2+\dot\theta_0^2)|{\vec\gamma}|^2
-{1\over2}\Lambda^4(T)\varepsilon^2\,\,\,\,\,\,\,\,\,\,\nonumber\\
&&+{\dot\chi_0^2/f_a^2\over2(1-N_\chi\chi_0^2/f_a^2)^3}\Bigg{[}(1+N_\chi\chi_0^2/f_a^2)^2(\sum_j\gamma_j)^2
\nonumber\\
&&\hspace{3cm}+N_\chi^2\chi_0^2/f_a^2(1-N_\chi\chi_0^2/f_a^2)|{\vec\gamma} |^2\Bigg{]}\nonumber\\
&&+{\chi_0\dot\chi_0\over(f_a^2-N_\chi\chi_0^2)^2}\Bigg{[}N_\chi(f_a^2-N_\chi\chi_0^2){\vec\gamma}\cdot\dot{\vec\gamma}\nonumber\\
&&\hspace{2.6cm}+(f_a^2+N_\chi\chi_0^2)(\sum_{i,j}\gamma_i\dot\gamma_j)\Bigg{]}
\eea

While this action looks somewhat complicated, things are significantly simplified by identifying the normal modes of the system, which we can decompose into an adiabatic mode and a collection of isocurvature modes.

\subsection{Adiabatic Mode}

\begin{figure}[t!]
    \centering
    \includegraphics[width=\linewidth]{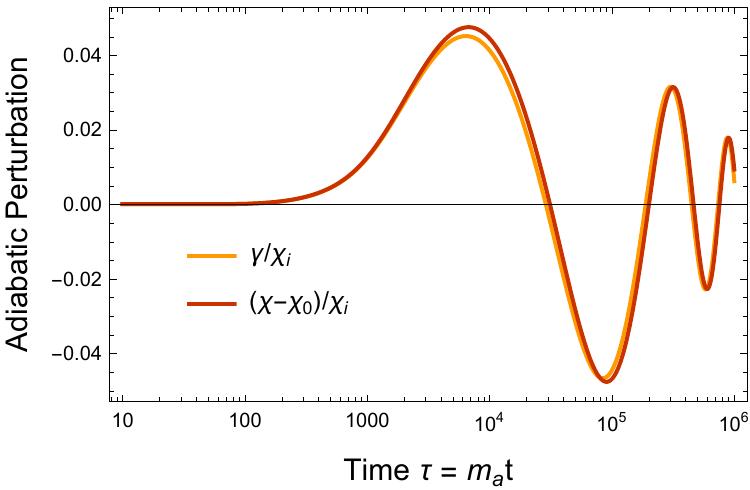}
     \caption{A homogeneous adiabatic perturbation in the $\chi_j$ fields by $\chi_j=\chi_0+\gamma$ for $\f\sim 10^{14}$ GeV and $m_\chi/ m_a = 10^{-5}$. We take $N_\chi=1000$ and the initial perturbation value $\gamma_i=10^{-4}\chi_i$ with $\chi_i=f_a/\sqrt{N_\chi+2}$. Note this $\gamma_i$ is only 1 order of magnitude smaller than the maximum value allowed for an adiabatic perturbation to be well defined given the constraint $\rho^2+\sum_j\chi_j^2=f_a^2$. The yellow curve is the result from the linear theory, while the red curve is from solving the full equations.}
    \label{fig:gamma}
\end{figure}

The adiabatic mode has 
\beq
\gamma\equiv\gamma_1=\gamma_2=\ldots=\gamma_{N_\chi}\,\,\,\mbox{and}\,\,\,\varepsilon\neq 0
\eeq 
The corresponding equations of motion in this case are
\bea
&&\ddot\gamma-{\nabla^2\over a^2}\gamma+\left(3H+{2N_\chi\chi_0\dot\chi_0\over f_a^2-N_\chi\chi_0^2}\right)\dot{\gamma}\nonumber\\
&&+\left((1-3N_\chi\chi_0^2/f_a^2)(m_\chi^2+\dot\theta_0^2)+{N_\chi(f_a^2+N_\chi\chi_0^2)\dot\chi_0^2\over(f_a^2-N_\chi\chi_0^2)^2}\right)\gamma\nonumber\\
&&\hspace{2cm}
+2\chi_0(1-N_\chi\chi_0^2/f_a^2)\dot\theta_0\,\dot\varepsilon=0\\
&&\ddot\varepsilon-{\nabla^2\over a^2}\varepsilon+\left(3H-{2N_\chi\chi_0\dot\chi_0\over f_a^2-N_\chi\chi_0^2}\right)\dot\varepsilon+{\Lambda^4(T)\over f_a^2-N_\chi \chi_0^2}\varepsilon\nonumber\\
&&\hspace{2cm}
-{2N_\chi\partial_t(a^3\,\chi_0\,\dot\theta_0\,\gamma)\over a^3(f_a^2-N_\chi\chi_0^2)}=0
\eea
In the equation for $\gamma$, we see that at early times when $\chi_0$ takes its initial value of $\chi_i=f_a/\sqrt{N_x+2}$, the coefficient of $(m_\chi^2+\dot\theta_0^2)\gamma$ is negative. This represents a tachyonic instability. The instability is tempered by the presence of Hubble friction and the other effective mass term, however there is still an instability. This is seen in fig.~\ref{fig:gamma}. This affects the {\em homogeneous} mode for $\gamma$ the most, as any derivatives only enhance the mass by $k^2/a^2$. Hence this means that the system may shift slightly to a different homogeneous mode. Plausibly, this instability will lead to a small shift in the final relic abundance. This has been verified numerically in the case worked out in fig.~\ref{fig:gamma}, but this deserves further consideration. 

\subsection{Isocurvature Modes}

For new qualitative behavior, we need to explore the remaining set of possible perturbations. 
The remaining $N_\chi-1$ eigenmodes are all of the isocurvature form in which every fluctuation in a $\gamma_j$ is compensated by an equal and opposite fluctuation in another $\gamma_{i\neq j}$, i.e.,
\beq
\sum_j\gamma_j=0\,\,\,\mbox{and}\,\,\,\varepsilon=0
\eeq
As an example, we can have
\beq
\gamma\equiv\gamma_1=-\gamma_2\,\,\,\mbox{and}\,\,\,\gamma_3=\cdots=\gamma_{N_\chi}=\varepsilon=0
\eeq
The equation for this $\gamma$ is of the relatively simple form
\bea
&& \ddot\gamma-{\nabla^2\over a^2}\gamma+3H\dot{\gamma}\nonumber\\
&&+\left(\left(1-{N_\chi\chi_0^2\over f_a^2}\right)(m_\chi^2+\dot\theta_0^2)+{N_\chi\dot\chi_0^2\over f_a^2-N_\chi\chi_0^2}\right)\gamma=0\,\,\,\,\,\,\,\,\,\,\,\,
\label{IsoEq}
\eea
Here we see that there is no tachyonic instability. Here the coefficient of $m_\chi^2+\dot\theta_0^2$ is the positive factor
\beq
\alpha\equiv 1-{N_\chi\chi_0^2\over f_a^2}
\eeq
At early times with $\chi_0=\chi_i=f_a/\sqrt{N_\chi+2}$ this takes on the small value $\alpha=1/(1+N_\chi/2)$. At later times, once $\chi_0$ has rolled down, then one has $\alpha\approx 1$. 

Since there is no tachyonic instability here, the homogenous mode is essentially stable. However, as $\theta_0$ oscillates there could be resonance into $\gamma$ modes of finite wavenumber. We now turn to study this possibility.

\section{Resonance}\label{Resonance}

\subsection{Parametric Resonance}

The driving term $\dot\theta_0^2\,\gamma$ can potentially give rise to parametric resonance. As a starting point, let us ignore Hubble expansion for the moment (we will compare to it soon). Once the axion is oscillating, we can approximate its background behavior as 
\beq
\theta_0=\theta_a\,\sin(\meff t)
\eeq
where the axion's effective mass $\meff=m_a/\sqrt{\alpha}$ is controlled by $\alpha$. The amplitude of oscillation $\theta_a$ is initially $\theta_a\sim 1$, and we shall treat its decreasing over time adiabatically. 

With $\chi_0$ either frozen at $\chi_i$ at early times or negligible at late times, the $k$-space isocurvature mode equation (\ref{IsoEq}) is
\beq
\ddot{\tilde{\gamma}}+(m_\chi^2\alpha+k^2+m_a^2\theta_a^2/2+m_a^2\theta_a^2\cos(2\meff t)/2)\tilde{\gamma}=0
\eeq
This is of the form of the Mathieu equation. For $\theta_a <1$ and $m_\chi\ll m_a$, we can focus on the narrow resonance regime, in which the maximum Floquet exponent can be shown to be given by (e.g., see Ref.~\cite{Hertzberg:2014jza})
\beq
\mu_{max}= {m_a^2\theta_a^2\over 8\,\meff}
\eeq
near the wavenumber $k_{max}=\sqrt{\meff^2-m_\chi^2\alpha-m_a^2\theta_a^2/2}$. 

Now we would like to include expansion in a simple adiabatic way. Here things appear complicated since the amplitude of oscillations $\theta_a$ is decreasing, while the axion mass increases due to its temperature dependence, until well after the QCD phase transition. However, there is a nice simplification: the axion number density is $n_a={1\over 2}\meff \feff^2\theta_a^2$ and due to entropy conservation, it red-shifts in the usual way as $n_a\propto 1/a^3$. Since $\meff \feff^2\propto\sqrt{\alpha}$ this implies that $\mu\propto 1/a^3$ also. 
We can compare this to Hubble; in a radiation era we have $H\propto 1/a^2$. Recall that oscillations begin when $3H\approx \meff$. So the ratio of the Floquet exponent to Hubble can be expressed over time as
\beq
{\mu_{max}\over H}\approx \alpha{3\theta_i^2\over 8}{a_{osc}\over a}
\eeq
where $a_{osc}$ is the scale factor when the axion starts oscillating. This means that when oscillations begin and $\theta_i\sim 1$, the ratio $\mu_{max}/H\sim \alpha_i=1/(1+ N_\chi/2)$. So for large $N_\chi$ the condition for resonance $\mu_{max}/H\gtrsim 1$ is not satisfied. At later times, when $\chi_0$ has itself decreased and $\alpha\approx 1$, the ratio $a_{osc}/a$ is small and the resonance condition $\mu_{max}/H\gtrsim 1$ is still not satisfied.

\subsection{Forced Resonance}

By causality, fields will tend to be inhomogeneous on super-horizon scales. As modes enter the horizon, they will acquire a wavelength of the order Hubble momentarily. This provides a type of inhomogeneous background of $\chi$ waves $\chi_s=\chi_s(t,{\bf x})$. These can be inserted back into the $\chi_j$ equation to act as a source for itself 
\beq
\chi_j=\chi_s+\cor_j
\eeq
By subtracting out the static piece of the axion $\langle\dot\theta_0^2\rangle$ to remove possible secular growth, we have
\beq
\ddot\cor_j-{\nabla^2\over a^2}\cor_j+3H\dot\cor_j+\tilde{m}_\chi^2\cor_j=-\alpha(\dot\theta_0^2-\langle\dot\theta_0^2\rangle)\chi_s
\eeq
where $\tilde{m}_\chi^2\equiv m_\chi^2\alpha+\alpha\langle\dot\theta_0^2\rangle$. 
Let us again ignore expansion and write the axion as $\theta_0=\theta_a\,\sin(\meff t)$.
Then we have
\beq
\ddot\cor_j-\nabla^2\cor_j+\tilde{m}_\chi^2\cor_j=-\theta_a^2m_a^2\cos(2\meff t)\chi_s/2
\eeq
We can expand the waves in terms of its Fourier transform as
\beq
\chi_s=\int{d^3k\over(2\pi)^3} (c_k\,e^{-i\omega_kt}+c_{-k}^*e^{i\omega_kt})e^{ikx}
\eeq
with $\omega_k=\sqrt{\tilde{m}_\chi^2+k^2}$. For $\omega_k\approx \meff$ this gives rise to resonant behavior from a forced oscillator. 
By imposing the initial condition $\cor_j(0)=\cor_j'(0)=0$, we can readily solve this equation. By passing to Fourier space and focussing on the near resonance wavenumbers we find
\beq
\tilde{\cor}_j ={i\,m_a^2\theta_a^2(c_k e^{i \meff t}-c_{-k}^*e^{-i \meff t})\sin(t(\meff-\omega_k))\over 8\meff(\meff-\omega_k)}
\eeq

Now the energy can be written as an integral over a kind of $k$-space density as
\beq
E_\cor=\int\! {d^3k\over(2\pi)^3}\,\rho_k
\eeq 
with
\beq
\rho_k=\sum_j\left({1\over2}|\dot{\tilde{\cor}}_j|^2+{1\over2}\omega_k^2|\tilde\cor_j|^2\right)
\eeq
By inserting the above solution, and again staying near resonance, we have
\beq
\rho_k=\sum_j{(|c_k|^2+|c_{-k}|^2)m_a^4\theta_a^4\sin^2(t(\meff-\omega_k))\over 64(\meff-\omega_k)^2}
\eeq
At late times we can simplify the time dependence by using the identity
\beq
{\sin^2(t(\meff-\omega_k))\over (\meff-\omega_k)^2}\to \pi\,t\,\delta(\meff-\omega_k)
\eeq
which is the standard simplification that leads to Fermi's golden rule.  This shows that the energy is growing linearly in time, as anticipated from a forced oscillator near resonance. We can insert into the energy integral, immediately carry out the radial integral using the delta function, giving the rate of change of energy into $\cor$ as
\beq
\dot{E}_\cor =\sum_j{V(\mathcal{N}_{k_r}+\mathcal{N}_{-k_r})m_a^4\theta_a^4 k_r\over 256\pi}
\eeq
where $k_r=\sqrt{\meff^2-\tilde{m}_\chi^2}$ is again the resonant wavenumber and 
 $V$ is the volume of some large box in which we perform this computation.
Here we have defined the occupancy number
\beq
\mathcal{N}_{k_r}\equiv {2|c_{k_r}|^2\,\omega_{k_r}\over V}
\eeq
It can be readily checked that the total number of particles of each species is given by
\beq
N_j=\sum_k\mathcal{N}_k = V\!\int\!{d^3k\over(2\pi)^3}\,\mathcal{N}_k
\eeq
so this definition of $\mathcal{N}_k$ makes sense. 

Now we are in a regime in which the energy density of axions is dominant and approximated as
\beq
\rho_a={1\over 2}m_a^2 f_a^2\theta_a^2
\eeq
with total energy $E_a=\rho_a\,V$. By energy conservation, the energy that is going into $\gamma$ must come from the axion; so we have $\dot{E}_a=-\dot{E}_\cor$. The corresponding relative rate of change of energy is
\beq
\Gamma_a={|\dot{E}_a|\over E_a}=\sum_j{(\mathcal{N}_{k_r}+\mathcal{N}_{-k_r})m_a^2\theta_a^2 k_r\over 128\pi\,f_a^2}
\eeq

On the other hand, we can compare this to the perturbative annihilation rate of 2 axions into 2 $\chi$ particles via the quartic coupling $\Delta\mathcal{L}=-|{\vec \chi}|^2(\partial\phi)^2/(2\feff^2)$. For non-relativistic axions, this can be readily shown to be
\beq
\Gamma_{2\to2}=n_a \sum_j\sigma_j v = n_a\sum_j {\meff \hat{k}_r\over 16\pi\,\feff^4}
\eeq
where $\hat{k}_r=\sqrt{\meff^2-m_\chi^2}$ is the on-shell wavenumber in vacuum.

Then noting that the axion number density is
\beq
n_a={1\over 2}\meff \feff^2\theta_a^2
\eeq
we see that the energy rate of change is (ignoring the tiny difference between $k_r$ and $\hat{k}_r$)
\beq
\Gamma_a\sim\alpha^2\sum_j(\mathcal{N}_{k_r}+\mathcal{N}_{-k_r})n_a\sigma_j v
\eeq
So we get the standard annihilation rate, but enhanced in the classical field regime by the $\chi_j$ occupancy numbers of the resonant modes $\mathcal{N}_{k_r}$ and $\mathcal{N}_{-k_r}$ (reflecting the fact that the particles are pair produced back to back) and suppressed by the factor $\alpha^2$ due to the dynamics being non-canonical. 

At the time at which the axion begins oscillating $\meff\sim 3H$, one can anticipate the relevant $\chi$ modes have typically wavenumber $k\sim H$ due to causality; this means they are near resonance. 
The occupancy is then $\mathcal{N}_{k_r}\sim \chi_s^2/k_{r}^2\sim \feff^2/\meff^2=\alpha^2 f_a^2/m_a^2$.
This gives an initial rate
\beq
\Gamma_{a,i}\sim \alpha^2N_\chi\, \meff\,\theta_i^2
\eeq
Since $\alpha^2 N_\chi\sim 1/N_\chi$ for large $N_\chi$, then this is much smaller than Hubble $H\sim\meff$ at this time.

Furthermore, this redshifts very quickly, because not only is there a factor of $1/a^3$ from the axion's number density. But also the resonant occupancy numbers are depleting due to redshifting. At later times, it is difficult to find $\chi$ waves with high occupancy at the axion's mass. 

\bibliography{DynPQ2}

\end{document}